\documentclass[12pt]{article}
\usepackage{amsmath, amssymb}

\usepackage{amsthm}

\usepackage{amsmath,amsthm}
\usepackage{multirow}
\usepackage{graphicx}
\usepackage{epstopdf}

\usepackage{ifthen}
\usepackage{color}

\ifthenelse{\isundefined{\GeneratePdf}}
{
}
{
\usepackage[pdftex]{graphicx}
\usepackage[pdftex]{hyperref}
}

\usepackage{framed}
\ifthenelse{\isundefined{\NoGenerateLetterSize}}
{

\ifthenelse{\isundefined{\GeneratePdf}}
{}
{
\pdfpageheight\paperheight
\pdfpagewidth\paperwidth
}
}
{}

\long\def\LongVersion#1\LongVersionEnd{}
\long\def\ShortVersion#1\ShortVersionEnd{#1}

\newcommand{\Comment}[1]{}

\LongVersion 
\LongVersionEnd 

\ShortVersion 
\ShortVersionEnd 

\ShortVersion 
\ShortVersionEnd 

\newtheorem{theorem}{Theorem}[section]
\newtheorem{lemma}[theorem]{Lemma}
\newtheorem{observation}[theorem]{Observation}
\newtheorem{corollary}[theorem]{Corollary}
\newtheorem{proposition}[theorem]{Proposition}
\newtheorem{claim}[theorem]{Claim}

\theoremstyle{definition}
\newtheorem{property}[theorem]{Property}
\newtheorem{definition}[theorem]{Definition}
\theoremstyle{plain}

\newtheorem{fact}[theorem]{Fact}
\newtheorem{example}[theorem]{Example}

\theoremstyle{definition}

\theoremstyle{plain}

\Comment{
\newtheorem{theorem}{Theorem}[section]
\newtheorem{lemma}{Lemma}[section]

\newtheorem{claim}{Claim}[section]
\newtheorem{proposition}{Proposition}[section]
\newtheorem{definition}{Definition}[section]

\theoremstyle{definition}

\theoremstyle{plain}
}

\newcommand{\Probability}[0]{\ensuremath{\hbox{\rm I\kern-2pt P}}}

\ShortVersion 
\renewcommand{\paragraph}[1]{\par\noindent\textbf{#1}}
\ShortVersionEnd 

\newcommand{\ignore}[1]{}

\begin{document}
\title{Recommendation Systems and Self Motivated Users}
\author{
Gal Bahar\thanks{
 Technion--Israel Institute of Technology, gal-bahar@campus.technion.ac.il}
 \and Rann Smorodinsky\thanks{
 Technion--Israel Institute of Technology, rann@ie.technion.ac.il. Smorodinsky gratefully acknowledges the support the United States-Israel Binational Science Foundation, National Science Foundation grant 2016734, German-Israel Foundation grant I-1419-118.4/2017, Ministry of Science and Technology grant 19400214, Technion VPR grants, and the Bernard M. Gordon Center for Systems Engineering at the Technion}
 \and Moshe Tennenholtz\thanks{
 Technion--Israel Institute of Technology, moshet@ie.technion.ac.il. Gal Bahar and Moshe Tennenholtz received funding from the Research Council (ERC) under the European Union’s Horizon 2020 research and innovation programme (grant agreement no. 740435).} }
\maketitle

\begin{abstract}
Modern recommendation systems rely on the wisdom of the crowd to learn the optimal course of action. This induces an inherent mis-alignment of incentives between the system's objective to learn (explore) and the individual users' objective to take the contemporaneous optimal action (exploit). The design of such systems must account for this and also for additional information available to the users. A prominent, yet simple, example is when agents arrive sequentially and each agent observes the action and reward of his predecessor. We provide an incentive compatible and asymptotically optimal mechanism for that setting. The complexity of the mechanism suggests that the design of such systems for general settings is a challenging task.
\end{abstract}

\section{Introduction}
Modern on-line recommendation systems (hereinafter called `mediators') have a two way relationship with their users (hereinafter called `agents'). Obviously they share information with agents by means of recommendations. However, they also turn to agents to collect information, which later on is processed into the aforementioned recommendations. It is convenient to think of this in the context of an abstract model that is inspired by the classical multi-armed bandit model. Self motivated agents arrive sequentially. At each stage the new agent is recommended which arm to pull. The agent is not confined to choose the recommended arm and may decide to choose another arm, based on some other information he may have. Whichever arm he pulls as well as the associated reward is observed by the mediator. This new information will later on be compiled into recommendations made to the following agents.

The goal of the mediator is to learn which arm performs better and only then to steer newly arriving agents to the optimal arm. However, agents are short lived and would like to exploit the information. Some concrete applications of this abstract model that are commercially popular are hotel choice (e.g., Agoda.com), travel routes (e.g., Waze and Google maps), service providers (e.g., Uber), out-sourcing freelancers (upwork.com), lending opportunities (e.g., Lendingclub.com) and so on and so forth. Note that in each of these settings there is a tension between the individual user who wants to take the optimal action and the objective of the mediator that wants to constantly learn the quality of new options (how good is the new hotel? How fast is a previously congested route? How proficient is the new UX developer?).

Using a distributed variant of multi-armed bandit to study the tension between the long term objective of the mediator and the myopic agents was first introduced in \cite{Kremer2014}. Their primary contribution is to design a mediator, a mapping from the information accumulated into recommendations at each stage, where it is in the best interest of agents to adopt the recommendations (`incentive compatibility'), yet in the long term the optimal arm will be identified. This exciting news has been extended in \cite{Mansour2015} and \cite{MansourSSW16} to several more elaborate bandit settings and to additional optimization criteria such as regret minimization.

The aforementioned papers make the implicit assumption that agents cannot see nor communicate with each other. This assumption may be unrealistic in some settings. When choosing a hotel via a mediator agents account for the recommendation but also account for what acquaintances recommend. In a restaurant example, agents will account for the choice chosen by their predecessors and tend to go to more crowded places. Unfortunately, the results obtained in \cite{Kremer2014}, \cite{Mansour2015} and \cite{MansourSSW16} are not robust to changes in the implicit assumption of no visibility. In fact, even with very little observability, for example when each agent just sees the action chosen by his immediate predecessor, the schemes proposed in those papers cease to be incentive compatible and lead to market failure.

The challenge of approximately optimal incentive compatible schemes subject to partial observability among the agents was addressed in \cite{BaharST16}. In that paper it is assumed that agents can view actions chosen by some of their predecessors but cannot observe the reward. They characterize structural properties that allow for the design of optimal schemes and provide explicit constructions.
In the current paper we enhance the observability assumption - not only do agents see the actions of predecessors but they can also see the reward. It turns out that whenever agents observe each other actions and rewards it is quite challenging for the mediator to get new agents to explore unknown or previously bad possibilities. Due to this challenge we focus on what is arguably a very elementary model, both in terms of reward distribution as well as of observability:
\begin{itemize}
\item
 We study a model with two arms only. The rewards distribution from one arm, which we refer to as the `safe' arm, are well known to all agents (as agents are risk neutral we can reduce the distribution to its expectation). The other arm, the `risky' arm, follows one of the two possible distributions, both with the same binary support.
\item
 Each agent only observes the action and reward of his immediate predecessor.
\end{itemize}
Our main contribution is the design of a mediator that is incentive compatible as well as approximately asymptotically optimal. That is, rational agents follow the meiator's recommendations while the optimal arm is eventually uncovered with high probability and thereafter all agents are recommended that arm.
We follow occum's razor and present a model with elementary settings and show that even in those settings, the mediator turns out to be quite complex. The resulting mediator is very different from those proposed in \cite{Kremer2014}, \cite{Mansour2015}, \cite{MansourSSW16} and \cite{BaharST16}.

An easy way to overcome the tension between the objectives of the mediator and those of the agents is by introducing payoffs. In this case the mediator can pay agents to explore a-priori inferior options. This could be quite costly and we would like to avoid such expenses. The mediator we propose does use payoffs. However it uses an arbitrarily small budget. In other words, for any positive budget, no matter how small, we propose a mediator that is incentive compatible (IC) and nearly optimal while being budget-balanced. As the budget shrinks to zero the corresponding mediator requires a bigger population of agents to reach approximate optimality.

The aforementioned literature can be viewed as extensions of the celebrated multi-armed bandit problems (\cite{BubeckC12}) to deal with setting where exploration can only be done by self motivated myopic agents and a central planner must incentivize exploration. Although our proposed mediator uses payments, the total budget spent can be arbitrarily small. Hence, our work better fits the above literature where payments are avoided, rather than work on incentevizing exploration through payments (e.g., \cite{FrazierKKK14}).

While the most related work to our study is the work on incentive-compatible exploration, work on data aggregation from strategic players can be found also in other ML contexts as described in e.g. \cite{CaiDP15}, \cite{DekelFP10}, \cite{LiuC16} and \cite{ShahZ15}.

\section{Model}
Let ${\mathcal N} = \{1,2, \ldots , N\}$  be the set of agents. Each agent chooses one of two actions from the set $A=\{R,S\}$ (hereinafter referred to as the risky arm and safe arm correspondingly). The reward from arm $S$ is fixed and commonly known to be $b$ while the reward from arm $R$ depends on the state of nature.%
\footnote{The model and results easily extend to the case where the reward over arm $S$ is random but with a commonly known expectation.}
Let $\Omega = \{H,L\}$ be the two possible states of nature, with prior probability $0 < q=prob(H) < 1$. Let $t^i \in \{0,1\}$ the (random) reward from arm $R$ at stage $i$. The sequence of random variables, $\{t^i\}_{i=1}^N$ are IID conditional on the state. Let $p_H = p(t^i = 1 | H)$ be the probability for a reward of $1$ at state $H$ and let $p_L = p(t^i =1 | L)$ be its counterpart in state $L$.

We make two assumptions on the model parameters:
\begin{itemize}
\item The risky arm may be superior, in expectation, or inferior to the safe arm, depending on the state of nature: $p_H>b>p_L$
\item The risky arm is a-priori superior to the safe arm:
    $q p_H + (1-q) p_L > b$. Clearly, without this assumption there is no way to convince an agent to try out the risky arm.
\end{itemize}

Agents arrive in a random order, $\sigma:N \to N$ (we abuse notation and use $N$ to denote the set of agents as well as the number of agents), where $\sigma$  is a random permutation chosen uniformly from all permutations. Let $i$ denote an agent and $j$ denote a stage. The equation $j=\sigma(i)$ suggest that agent $i$ arrived at stage $j$. In what follows we mostly refer to agents according to the (random) stage at which they arrive and not according to their name. We denote by $a^j$ is the action of the agent arriving at stage $j$ (and not the action of agent $j$). The reward of the agent arriving at stage $j$, denoted $r^j$, satisfies $r^j = t^j$ whenever $a^j = R$ and $r^j=b$ whenever $a^j = S$.

Each agent observes:
\begin {itemize}
\item Which arm his predecessor pulled.
\item The reward of his predecessor
\item some private message sent to him by the mediator (the agent does not observe the message of the mediator to his predecessor).
\end{itemize}
 The input for the mediator, at stage $j$, is the sequence of rewards (which subsumes the information about the chosen arms) received by the agents at all previous stages ($\{0,1,b\}^{j-1}$). The output of the mediator is a message $m$ in some abstract set $M$. Formally, a mediator is a function
$mdr:\cup_{j=1}^N  \{0,1,b\}^{j-1} \rightarrow M$. Without loss of generality we can assume that the message space, $M$, has the form $M=A \times \hat M$, which means that the recommendation for an arm to pull in $A$ is only a part of the message.

We extend the notion of a mediator in two directions:
\begin{itemize}
\item First we allow for a random message. To capture this we introduce an abstract input  space, $\Phi$, and allow the message to depend on the random realization of $\phi \in \Phi$.
\item In addition we extend the setting and allow the mediator to make payments (in the same currency as the rewards). Therefore a mediator is random variable $\phi \in \Phi$ and a function
$mdr:\cup_{j=1}^N \{\Phi \times \{0,1,b\}^{j-1}\} \rightarrow A \times M \times \mathcal{R}_+$, which includes, inter-alia, a recommended action as well as a subsidy that is paid to the agents whenever this recommendation is accepted. We denote by $Im(mdr)$ the image of the mediator.
\end {itemize}

Formally, the utility of an agent $i$ is $u^i(a^{\sigma(i)},m^{\sigma(i)},\omega)$ and it is a function of the agent's action $a^{\sigma(i)}$, the message $m^{\sigma(i)}=(m^{\sigma(i)}_a,m^{\sigma(i)}_s,m^{\sigma(i)}_p)$ and the state of nature, $\omega$, as follows:
         \begin{itemize}
            \item If $a^i = S = m^{\sigma(i)}_a$ then $u^i = b + m^{\sigma(i)}_p$
            \item If $a^i = S \neq m^{\sigma(i)}_a$ then $u^i = b$
            \item If $a^i = R = m^{\sigma(i)}_a$ then $u^i = t^{\sigma(i)} + m^{\sigma(i)}_p$
            \item If $a^i = R \neq m^{\sigma(i)}_a$  then $u^i = t^{\sigma(i)}$
         \end{itemize}

Using large subsidies the mediator can induce agents to act in any way it pleases by paying them enough to follow his recommendation. However, our challenge is to construct a `lean' mediator, one that uses a limited budget (arbitrarily small) and yet learns the state of nature and steers most agents toward the optimal action.

A strategy for agent $i$ is a function $\rho^i: \{0,1,b\} \times M \rightarrow A$ which determines which arm to be pulled by agent $i$  as a function of   $r^{\sigma(i)-1}$ and $m^{\sigma(i)}$. It is called {\em Dominant} if
$u^i(\rho^i(r,(m_a,m_s,m_p)), (m_a,m_s,m_p) ) \ge u^i(a,(m_a,m_s,m_p))$ for any $r\in{0,1,b}$, any message $(m_a,m_s,m_p) \in  Im(mdr)$ and any $a \not=\rho^i(r,(m_a,m_s,m_p))$.

Recall that a generic message is a triplet $m=(m_a,m_s,m_p)$. Agent $i$'s strategy is {\em mediator-respectful} if $a^{\sigma(i)}=m^{\sigma(i)}_a$. In words, if he pull the arm recommended by the mediator no matter which arm he observed his predecessor pulled and what was the reward he observed his predecessor got. If the mediator-respectful strategy is dominant for all agents, $i$, then the mediator is said to be {\em Incentive-Compatible}. %
\footnote{As a tie-break we assume that if the agent is indifferent between the two arms he will pull the arm that is recommended by the mediator}

More formally, $mdr$ is {\bf Incentive Compatible (IC)} if $\forall 1\leq i \leq N$, for any history
$h=(\phi ;h_1,\ldots,h_{i-1})$ in $\Phi \times \{0, 1, b\}^{i-1}$
and for any action $a \in A$:
$$E((mdr(h))_a | mdr(h), h_{i-1}) + (mdr(h))_p  \geq  E(a|mdr(h), h_{i-1}).%
\footnote{We pursue the terminology of \cite{MyersonJME1982} and use the notion of incentive comptability for mediators that communicate an action, $a \in A$, instead of some abstract message, to each agent. Such a mediator is `incentive compatible' if each agent finds it in her best interest to pursue the recommended action.}
$$

Our goal is to provide a mediator that induce sufficiently many agents to explore the risky arm and use a negligible budget to do so.

A mediator is $\beta$-{\bf budget balanced} if for any realization of $\phi$ and the sequence $t_i$, $\Sigma_{i=1}^N m^i_p \leq \beta$. In other words, the sum of the subsidies promised never surpasses $\beta$.
A mediator is $\epsilon$-{\bf optimal} if it is incentive compatible and
$\frac{\Sigma_{i=1}^N r^i}{N} \geq (1-\epsilon)p_H$ in state $H$ and
$\frac{\Sigma_{i=1}^N r^i}{N} \geq (1-\epsilon)b$ in state $L$.

\section{The mediator}

Assume one knows how to construct a mediator when agents do not observe each other. In our setting each agent observes only his immediate predecessor and so agents in odd stages do not observe each other. Thus, an intuitive extension of the preliminary mediator is to implement the preliminary mediator only on odd stages (while ignoring the input from even stages). Unfortunately, this will not work as information may be passed on indirectly. To see why consider the agent at stage $3$, denoted $g_3$. If she observed $g_2$ pulling the safe arm she can conclude that $g_2$ observed $g_1$ pulling the safe arm or, otherwise, that $g_1$ pulled the risky arm but obtained a low reward ($L$). Hence, $g_3$ (and similarly $g_5$, $g_7$ and so) may not comply with a recommendation to check out the risky arm. In short, IC is not guaranteed.

This failed example sheds some light on the inherent difficulties to design a mediator:
\begin{itemize}
   \item Agents are self motivated and will not agree to pull the recommended arm unless they think it is in their best interest to do so.
   \item The mediator can only learn by observing a large enough sample of the risky arm. On the other hand, roughly speaking, agents that observe their predecessor pull the risky arm and get a low reward or observe the predecessor pull the safe arm will be inclined to pull the safe arm. Hence, the mediator must. somehow, persuade many such agents to pull the risky arm.
   \item By recommending agents to use the risky arm the mediator conveys additional infirmation implicitly. This, in turn, could influence their belief over which arm is optimal, which may no longer be the recommended one.
  \item The mediator can only use a negligible budget.
\end{itemize}

We construct a formal parametric family of mediators, which we dub the {\em Innkeeper's mediator}, for reasons we describe in the footnote.%
\footnote{An innkeeper recommends a local seafood restaurant to his customers. The quality of the restaurant is well known and the innkeepr is happy to recommend it. One day a new restaurant opens up, the quality of which is not known yet, and could possibly be better or worse than that of the familiar place. As the innkeeper has shellfish allergy he cannot check out the new restaurant and must rely on patrons’ feedback. Due to the inherent volatility in quality, feedback is required from a large sample. The inn has one room and each guest, upon checking in, meets the previous guest and in particular learns which restaurant he visited and whether he enjoyed it. The algorithm we design allows the innkeeper to learn the quality of the restaurant and eventually steer his customers to the better restaurant.}
We then argue that all these mediators are IC and in addition are asymptotically optimal and budget balanced.

\subsection{An informal description}

We provide the formal description of the innkeeper mediator in terms of a pseudo-code. However, before doing so we start with an informal description that provides some intuition as to why it is IC and asymptotically optimal.


We break down the description of the innkeeper mediator to its three main phases. In the first phase it only accumulates information and gains an informational advantage over the agents without providing any  recommendation. In the second phase the mediator use his informational advantage in order to acquire all the knowledge needed to evaluate the state of nature. In the third phase the mediator exploits this information.



\begin{itemize}
\item First phase - the `pre-intervention' phase: During the first phase the mediator does not interfere (formally, this is done by recommending to each agent what the agent would have done in the absence of the mediator). The length of this phase, $K$,  is set so that it is sufficient to deduce the true reward distribution of the risky arm, if all agents during this stage actually choose to pull it.
     The pre-intervention phase necessarily ends in one of the following options:
\begin{enumerate}
\item
All $K$ agents pulled the risky arm and many of them enjoyed a high reward. In this case the mediator can conclude that the state is very likely to be $H$.
\item
All $K$ agents pulled the risky arm and many of them were disappointed, in which case the mediator can conclude that the state is very likely to be $L$.
\item
Some of the $K$ agents pull the safe arm and so the mediator might lack substantial knowledge about the risky arm.
\end{enumerate}
Consider an arbitrary agent that arrived after time $K$. Assuming this agent only knows that case (2) or (3) have materialized (and nothing else) he would then prefer the safe arm. This statement is obvious for case (2). To see why this also holds in case (3) note that once an agent, among the first $K$, pulls the safe arm his successor will copy him. The reason is that the successor must think that the predecessor pulled the safe arm only because this was the optimal thing given the evidence available to him. With no additional information to the successor he will follow the lead.
 On the other hand, if he knows case (1) materialized then he would definitely pull the risky arm.

The mediator, in order to convince agents to pull the risky arm even if case (3) materializes, reveals this information partly in the following way. In case (1) he tosses a biased coin (with parameter $\delta$). Whenever the coin falls on Heads he announces that case (1) happened and that he moves to the exploit phase (see below). If the coin falls on Tails or if cases (2) or  (3) materialized he announce that he moves to the switching phase. The parameter $\delta$ can be chosen so that the aforementioned agent becomes indifferent between pulling the two arms, and might as well accept the recommendation.

\item Second phase - `switching' phase:
During the second phase we guarantee with high enough probability exactly $K$ switches in both directions - by switching we refer to a scenario where an agent pulls one arm and his predecessor pulls the other arm. This is done for two reasons:
\begin{enumerate}
\item It guarantees that $K$ agents will pull the risky arm and, by doing so, the mediator can learn the state of nature.
    \item Maintain the indifference between the two arms for any agent that is recommended a switch. This, in turn, will ensure that agents comply with the mediator's recommendation.
\end{enumerate}
The circumstances for which the mediator recommends a switch are as follows. Anytime an agent pulls the safe arm, the next agent will be recommended the risky arm. Also, anytime an agent pulls the risky arm and receives a low reward the mediator will recommend to switch to the safe arm.

 It turns out that with the correct choice of the coin's parameter, $\delta$, this recommendation will become incentive compatible.  Following $K$ switches in each direction ($2K$ switches in total) the mediator is ready to move to the exploitation stage.

 Unfortunately, the number of periods for these $2K$ switches is an unbounded random variable. However, by controlling the length of the switching phase to some large integer we can guarantee $2K$ switches will occur with an arbitrary high probability. The introduction of this bound distorts the aforementioned indifference between the two arms. The longer the phase is the smaller the distortion. To resurrect indifference  we introduce transfers. The total transfers (the budget) shrink to zero as the distortions become smaller and hence can be made arbitrarily low.

\item Third phase -`exploitation: In this phase the mediator collects no more information and consistently recommends, what he deems, is the best arm. This phase is triggered only if one of the following two events occurs:
\begin{enumerate}
 \item  Case (1) of the pre-intervention phase was materialized and the coin with probability $\delta$ was falling on Tail. However, in that case the mediator had $K$ samples of $R$ during the pre-intervention phase and it determined with high enough probability that the world state of $R$ is $H$.
 \item The switching phase was finished with $2K$ switches. In this case the mediator had at least $K$ samples of $R$ during the exploration phase and therefore can determine the world state of $R$ with high enough probability.
\end{enumerate}
\item Finally, to ensure agents follow the mediator even when some predecessor deviated, then any such deviant would trigger a no-intervention policy once again.
\end{itemize}


\subsection{Pseudo-code}

We begin by describing the input space, the message space and the parameters of the mediator and then provide a pseudo-code that details the mediator.

The random input $\phi$ is the result of a coin flip and hence it is in $\{0,1\}$.  Hereinafter we replace $\phi$ with $cf \in \{0,1\}$ as a random variable determined by a coin flip with probability $\delta$ for the outcome $1$.

The message space is $M =\{R,S\} \times \hat{M} \times \mathcal{R}_+$, with generic message $m=(m_a,m_s,m_p)$:
\begin{itemize}
\item  $m_a \in \{R,S\}$ is the action recommended by the mediator.
\item  $m_s \in \hat{M} = \{s_1, s_2, s_3, s_4\}$ is some extra information offered to the agent. This information encodes the phase at which the mediator is at. There are 4 phases:
    \begin{itemize}
    \item A pre-intervention phase, encoded $s_1$.  At this phase each agent is recommended the arm he would have preferred based on his own knowledge.
    \item A switching phase, encoded $s_2$. In this phase whenever an agent pulls the safe arm his successor is recommended the risky arm and whenever an agent pulls the risky arm and receives a low reward his successor is recommended the safe arm.
    \item An exploitation phase, encoded $s_3$.
    \item A deviation treatment phase, encoded $s_4$, is triggered whenever some agent did not follow the mediator's recommended action. If a deviation occur before the exploitation phase each agent will be recommended the arm he would have preferred based on his own knowledge. Deviations that occur during the exploitation phase are ignored.
\end{itemize}
\item $m_p$ is the subsidy.
\end{itemize}

Some auxiliary variables required for the pseudo-code are:

\begin{itemize}
\itemsep0em
\item Let $x_i = 1$ whenever $(r^{i-1}=0 \cap m^i_s = s_2)$ and $x_i = 0$ otherwise. Let $X=\Sigma_{i=1}^N x_i$. In words, $X$ counts the number of stages during the second phase (the switching phase) in which an agent observed his predecessor pull the risky arm and receive a low reward ($r^{i-1}=0$).
\item Define $R_1 = [\cap_{1 \leq i \leq K}\{a^i = R\}] \cap (\frac{\Sigma_{i=1}^Kt^i}{K} \geq \frac{p_H+p_L}{2})\}$. In words, $R_1$ is the event where all agents in stages $1,\ldots,K$ pulled the risky arm and their average reward exceeded $\frac{p_H+p_L}{2}$.
\item Let $R_2 = [\cap_{1 \leq i \leq K}\{a^i = R\}] \cap (\frac{\Sigma_{i=1}^Kt^i}{K} < \frac{p_H+p_L}{2})\}$. In words, $R_2$ is the event where all agents in stages $1,\ldots,K$ pulled the risky arm and their average reward fell below $\frac{p_H+p_L}{2}$.
\item Let $S = \cup_{1 \leq i \leq K} \{a^i = S \}$ be the event where some agent is stages $1,\ldots,K$ pulled the safe arm (we abuse notation and use the same notation, $S$, for the safe arm and for an event. The correct interpretation of $S$ is always obvious from the context).
\item Define $s_2 = ((R_1\cap cf=1)\cup R_2 \cup S)$. In the event $s_2$ the mediator will move to the switching phase and otherwise it will begin exploitation. Note that the complement of the event $s_2$ is the event that the random coin, $cf$, equals $0$ and in addition the initial (pre-intervention) phase ended with all agents pulling the risky arm and the conclusion is that this arm is most likely superior to the safe one.
\item Define $exploit\_flag$ as a binary variable receiving the value `true' whenever the  mediator in the exploitation phase.
\item Define $deviation\_flag$ as a binary variable receiving the value `true' whenever a deviation of an agent is detected (namely, an agent does not follow the recommended action).
\item Let $switching\_count$ count the number of agents, during the switching phase, who pulled a different arm from their predecessor.
\item  $R\_count$ counts the number of times the risky arm was pulled during the switching phase.
\item $R\_sum$ keeps track of the total rewards received from the risky arm during the switching phase.
\end{itemize}
For any quadruple of parameters, $N,\beta,\delta,K$, where $\delta$ is the coin parameter, $N$ the
population size, $K$ the sample size and $\beta$ the budget constraint, we define the  corresponding innkeeper mediator, denoted $INKP(N,\beta,\delta, K)$, as follows: (\textcolor{blue}{the clarifying comments in blue are not part of the pseudo-code})

\itemsep0em
\begin{center}
{\bf The innkeeper mediator $INKP( N,  \beta, \delta, K)$ }
\end{center}
\textcolor{blue}{ -------------------------- Initiation  ---------------------}
\begin{itemize}
\itemsep0em
\item Set $exploit\_flag = false$
\item Set $deviation\_flag = false$
\item Set $R\_sum = 0, R\_count=0$
\item Set $switching\_count = 0$
\end{itemize}
\textcolor{blue}{ ------------------- On the equilibrium path ------------}
\begin{enumerate}
\item While $deviation\_flag \neq true$ do:

\begin{itemize}
\item $\tilde M^{1} = (R,s_1,0)$ ; If $a^1 \neq R$ set $deviation\_flag = true$.

\item For $\sigma^{-1}(i): i= 2, \ldots,N$: \\
\textcolor{blue} { -------------------------- Pre intervention phase  ------------------}

\begin{itemize}
  \item While $i\leq K$ do:
 If $(E(t^i|r^{i-1}) > b)$ Set $m^i_a = R$ else set  $m^i_a = S$. \\
 $\tilde M^{i} = (m^i_a,s_1,0)$ ; If $a^i \neq m^i_a$ set $deviation\_flag = true$.
  \item If $i=K+1$:
  \begin{itemize}
  \item Flip a coin with probability $\delta$, and with probability $\delta$ set $cf=1$ else set $cf=0$
  \item  Check the following 3 conditions:
      \begin{enumerate}
          \item $\forall 1 \leq j \leq K: a^j = R$
          \item $\frac{\Sigma_{j=1}^Kr^j}{K} \geq \frac{p_H+p_L}{2}$
          \item $cf=0$ (Recall $cf$ is determined by a coin flip with probability $\delta$ see Lemma 3 in  appendix for $\delta$)
      \end{enumerate}
  \item If all 3 conditions are satisfied set $exploit\_flag = true, exploit\_value = R$ and for all $i>K$ set $\tilde M^{i} = (R,s_3,0)$ ; If $a^i \neq m^i_a$ set $deviation\_flag = true$.
  \item If not all 3 conditions are satisfied set $switching\_count=0$ and continue to the switching phase
  \end{itemize}
  \textcolor{blue} { -------------------------- Switching phase --------------------------------}\\
  \item while $switching\_count < 2K$ do:
  \begin{itemize}
  \item if $R\_count < K$ and $a^{i-1} = R$
     \begin{itemize}
         \item $R\_count = R\_count +1$
         \item $R\_sum = R\_sum + r^i$
     \end{itemize}
  \item If $r^{i-1} = 1$: set $\tilde M^{i} = (R,s_2,0)$; If $a^i \neq m^i_a$ set $deviation\_flag = true$.
  \item If $r^{i-1} = 0$: set $\tilde M^{i} = (S,s_2, \frac{\beta}{2K})$ and set $switching\_count = switching\_count + 1$; If $a^i \neq m^i_a$ set $deviation\_flag = true$.
 \item If $r^{i-1} = b$ set $\tilde M^{i} = (R,s_2, \frac{\beta}{2K})$ and set $switching\_count = switching\_count + 1$ ; If $a^i \neq m^i_a$ set $deviation\_flag = true$.
  \end{itemize}
\textcolor{blue} { -------------------------- Exploitation phase ----------------------------}
 \item If $switching\_count = 2K$ (note that $switching\_count = 2K$ ensure at least $K$ agents took action $a$ during the switching phase)
 \begin{itemize}
    \item If $\frac{R\_sum}{K}\geq \frac{p_H + p_L}{2}$ set $exploit\_flag = true, exploit\_value = $ and for all $i$ set $\tilde M^{i} = (R,s_3,0)$ ; If $a^i \neq m^i_a$ set $deviation\_flag = true$.
    \item If $\frac{R\_sum}{K}< \frac{p_H + p_L}{2}$ set $exploit\_flag = true, exploit\_value = S$ and for all $i$ set  $\tilde M^{i} = (S,s_3,0)$; If $a^i \neq m^i_a$ set $deviation\_flag = true$.
 \end{itemize}
\end{itemize}
\end{itemize}
\textcolor{blue} { ------------------------ Off the equilibrium path  ------------------}
\item while $deviation\_flag =true$ do
\begin{itemize}
\item if $exploit\_flag = True$: set  $\tilde M^{i} = (exploit\_value,s_3,0)$ for all $i$ \textcolor{blue}{* Ignore deviations that occur in the exploitation phase *}
\item if $exploit\_flag = false$: for all $i$ do:\\
 If $(E(t^i|a^{i-1}\cap r^{i-1} \cap m^i_s=s_4 ) > b)$ Set $m^i_a = R$ else set  $m^i_a = S$. \\
 $\tilde M^{i} = (m^i_a,s_4,0)$  \textcolor{blue}{* Deviations that occur before the exploitation phase was triggered - each agent is recommended the best response conditional on the agent's knowledge. *}
\end{itemize}
\end{enumerate}

\section{Main Result and Proof sketch}
We are finally ready to address our primary challenge. We show that for any pair of parameters, $\beta$ and $\epsilon$,  there exists some number $N$ and a corresponding incentive compatible, $\beta$-budget balanced and $\epsilon$-optimal mediator for any population with $N$ agents or more.

\begin{theorem}\label{THEOREM1}
For any $\epsilon, \beta>0$ there exist $N' \ge K>0$ and $\delta \in [0,1]$ such that the innkeeper mediator $INKP( N, \beta, \delta, K)$ is incentive compatible, $\beta$-budget balanced and $\epsilon$-optimal for all $N\geq N'\ $.
\end{theorem}

\noindent
 {\bf Sketch of proof:} A formal proof is provided in an on-line appendix. We now explain the main ingredients of this proof by expanding parts our informal description of the mediator (Section 3.1).
 
 In the pre-intervention phase the mediator is, de-facto, silent. In practice it submits a recommendation for an arm which the agent would have pulled in the hypothetical case that the mediator does not exist. The length of this phase, $K$,  is computed to ensure that a sample of size $K$ from the risky arm will reveal the state of nature with high probability (see Lemmas 1 and 2 in the appendix). Recall this phase ends up in one of the following cases:
\begin{enumerate}
\item
All $K$ agents pulled the risky arm and were a majority of them was rewarded $1$. In this case the mediator can conclude that the state is very likely to be $H$.
\item
All $K$ agents pulled the risky arm and were a majority of them was rewarded $0$. In this case the mediator can conclude that the state is very likely to be $L$.
\item
Some of the $K$ agents pulled the safe arm.
\end{enumerate}

While in cases (2) and (3) the mediator will continue to the switching phase, in case (1) it will be decided by coin flip with probability $\delta$ if the mediator will continue to the switching phase or jump strictly to the exploit phase. As a result, the mediator might know the better choice (with high probability) but will choose not to exploit just yet. This happens in case (1) and whenever the coin flip comes out Head or in case (2). This is done in order to allow for
exploration in case (3).

In Lemma 3 in the appendix we prove that there exists a $\delta \in [0,1]$ that will make a hypothetical `blind' agent in the switching phase indifferent between the two arms. A `blind' agent is an agent who does not observe his predecessor.

In reality agents are not blind and could observe one of three outcomes and recommendations during the switching phase:
\begin{enumerate}
\item
The predecessor pulled the risky arm and was rewarded $1$. The mediator recommends the risky arm again. In Lemma 4 in the appendix we prove that this recommendation is IC.
\item
The predecessor pulled the risky arm and was rewarded $0$. The mediator recommends to switch to the safe arm. In  part (1) of Lemma 8 in the appendix we prove that this recommendation is IC.
\item
The predecessor pulled the safe arm. In that case the mediator will recommend him to switch to the risky arm. In part (2) of Lemma 8 in the appendix  we prove that this recommendation is IC.
\end{enumerate}

The switching phase will go on until $2K$ switches occur, in which case the required information is accumulated and the mediator can turn to the exploitation stage. However, it may be the case that the game ends before these $2K$ switches occur and so no exploitation takes place. The possibility that the switching phase will not end introduces two problems:
\begin{itemize}
  \item First, there is some probability that not enough information is accumulated.
  \item Second, it introduces a distortion in the incentives and the IC property no longer holds.
\end{itemize}
    However, by extending the switching phase to be long enough we can bound from below the probability that such $2K$ switches occur arbitrarily close to $1$. In Lemma 7 in the appendix we introduce the desired length of the switching phase that guarantees that the $2K$ switches will indeed occur with high probability.  In Lemma 8 in the appendix we show that by introducing transfers, the magnitude of which is arbitrarily low, we can resurrect the IC property.

The proof of Theorem 1 is split to three propositions (see our on-line appendix): Proposition 1 proves incentive compatibility, Proposition 2 accounts for $\epsilon$-optimality and budget balancedness is shown in Proposition 3.

\section{Appendix: Proofs}

In this section we provide a sequence of lemmas and their proofs that are required for the proof of our main result. The lemmas and their proofs are followed by the proof of our main result, which split into 3 propositions.  

\subsection{Determining the sample size, $K$}

The following first two lemmas are instrumental in computing the sample size given the model parameters and the required level of efficiency:

\begin{lemma}\label{THM1}
For any $\epsilon > 0$ there exists $K=K(\epsilon)>0$ such that:
\begin{enumerate}
\item $p(\frac{\Sigma_{i=1}^{K}t^i}{K}\geq \frac{p_H+p_L}{2}|H)\geq (1-\epsilon)$
\item $p(\frac{\Sigma_{i=1}^{K}t^i}{K}  < \frac{p_H+p_L}{2}|L)\geq (1-\epsilon)$
\end{enumerate}
\end{lemma}

In fact, the required sample size is $K=\max\{K_H,K_L\}$, where  $K_H = \frac{4p_H(1-p_H)}{\epsilon(p_H-p_L)^2}$ and $K_{L} = \frac{4p_L(1-p_L)}{\epsilon(p_H-p_L)^2}$.

\textbf{Proof}:

Set $K_H = \frac{4p_H(1-p_H)}{\epsilon(p_H-p_L)^2}$, $K_{L} = \frac{4p_L(1-p_L)}{\epsilon(p_H-p_L)^2}$ and
$K = max(K_{H},K_{L})$. Let $X=\sum_{i=1}^K t^i $ be the sum of the $K$ rewards from pulling the risky arm.

Conditional on the state $H$ the expectation of $X$ is $E(X) = Kp_H$ and its variance is $Var(X) =Kp_H(1-p_H)$.

\begin{equation}\label{eqth11}
\begin{split}
&p(\frac{\Sigma_{i=1}^{K}t^i}{K}< \frac{p_H+p_L}{2}|H)
= Prob(X < K\frac{p_H+p_L}{2}) = Prob(E(X)- X > E(X)-K\frac{p_H+p_L}{2}) \le \\
&\le Prob(E(X)- X > E(X)-K\frac{p_H+p_L}{2})) + Prob(X-E(X) > E(X)-K\frac{p_H+p_L}{2}) = \\
&= Prob(|E(X)- X| > E(X)-K\frac{p_H+p_L}{2}).
\end{split}
\end{equation}

Applying Chebyshev's inequality:
\begin{equation}\label{eqth12}
\begin{split}
&Prob(|E(X)- X| > E(X)-K\frac{p_H+p_L}{2}) \le  \frac {Var(X)}{(K\frac{p_H+p_L}{2}-E(X))^{2}} =
\frac{Kp_H(1-p_H)}{(K\frac{p_H-p_L}{2})^2}\\
& = \frac{4p_H(1-p_H)}{K(p_H-p_L)^2}
\end{split}
\end{equation}

However, since $K \geq \frac{4p_H(1-p_H)}{\epsilon'(p_H-p_L)^2}$, from equations \ref{eqth11} and \ref{eqth12} we get
\begin{equation}\label{eqth13}
\begin{aligned}
p(\frac{\Sigma_{i=1}^{K}t^i}{K}< \frac{p_h+p_L}{2}|H)\leq \epsilon'
\end{aligned}
\end{equation}

The proof of the complementary case, when the state of nature is $L$, follows similar arguments and is therefore omitted.

QED  \\

\begin{lemma}\label{THM2}
There exists $K_b$ such that
$\forall K_b+1 \leq j \leq N$ the following inequality holds:
$E(t^j|\frac{\Sigma_{i=1}^{K_b}t^i}{K_b}< \frac{p_H+p_L}{2}) < b$
\end{lemma}

This is essentially a corollary of Lemma \ref{THM1}.

\textbf{Proof}:
\begin{equation}\label{eqt21}
\begin{split}
&   \forall K_b+1 \leq j \leq N: E(t^j|\frac{\Sigma_{i=1}^{K_b}t^i}{K_b}< \frac{p_H+p_L}{2})= \\
&   p(H|\frac{\Sigma_{i=1}^{K_b}t^i}{K_b}< \frac{p_H+p_L}{2})p_H + p(L|\frac{\Sigma_{i=1}^{K_b}t^i}{K_b}< \frac{p_H+p_L}{2})p_L \leq \\
&   \frac{p(H)}{p(\frac{\Sigma_{i=1}^{K_b}t^i}{K_b}< \frac{p_H+p_L}{2})}p(\frac{\Sigma_{i=1}^{K_b}t^i}{K_b}< \frac{p_H+p_L}{2})|H)+p_L  \leq \\
&   \frac{1}{p(\frac{\Sigma_{i=1}^{K_b}t^i}{K_b}< \frac{p_H+p_L}{2})}p(\frac{\Sigma_{i=1}^{K_b}t^i}{K_b}< \frac{p_H+p_L}{2})|H)+p_L
\end{split}
\end{equation}

Set $\epsilon' = min(0.5,\frac{(1-q)(b - p_L)}{2})$  and let $K=K(\epsilon')$ be the corresponding sample size from Lemma \ref{THM1}. We can now apply the inequalities in Lemma \ref{THM1} to conclude that:

\begin{equation}\label{3eqtc21}
\begin{split}
&   \forall K_b+1 \leq j \leq N: E(t^j|\frac{\Sigma_{i=1}^{K_b}t^i}{K_b}< \frac{p_H+p_L}{2}) \leq
\frac{1}{p(\frac{\Sigma_{i=1}^{K_b}t^i}{K_b}< \frac{p_H+p_L}{2})}\epsilon'+p_L =\\ &\frac{1}{qp(\frac{\Sigma_{i=1}^{K_b}t^i}{K_b}< \frac{p_H+p_L}{2}|H)+(1-q)p(\frac{\Sigma_{i=1}^{K_b}t^i}{K_b}< \frac{p_H+p_L}{2}|L)}\epsilon'+p_L \leq \\
& \frac{1}{(1-q)p(\frac{\Sigma_{i=1}^{K_b}t^i}{K_b}< \frac{p_H+p_L}{2}|L)}\epsilon'+p_L \leq  \frac{1}{(1-q)(1-\epsilon')}\epsilon'+p_L \leq b
\end{split}
\end{equation}

Q.E.D \\

\subsection{Exploit or switch?}

On our path to proving the main theorem we make some interim observations. The first one argues that there exists a parameter $\delta$ that induces indifference between the two arms once the switching phase begins (assuming no additional information is available):


\begin{lemma}\label{THMDELTA}
 For any $\epsilon, \beta >0$ there exist $K'=K(\epsilon)$ and $ \delta \in[0,1]$ such that for any  $N>K>K'$ and any $N \geq  i >K$, the innkeeper mediator $INKP(N, \beta, \delta, K)$ satisfies  $E(t^i|M^i_s=s_2)=b$.
\end{lemma}

The proof follow the mean value theorem. First we prove that in case $\delta = 1$ the risky arm will be preferable. Then we prove that in case $\delta = 0$ the safe arm will be preferable. Last we prove that the expected value of the safe arm is continues in $\delta \in [0,1]$ and the proof of the lemma will result from the mean value theorem.

\textbf{Proof}:

Recall that the event $\{m^i_s=s_2\}$ is equivalent to the event $\{(R_1\cap cf=1)\cup R_2 \cup S\}$. Therefore:
\begin{equation} \label{eqt171}
\begin{split}
&\forall K < i \leq N: E(t^i|M^i_s=s_2) = p(R_1| M^i_s=s_2)E(t^i|R_1\cap M^i_s=s_2) + \\
&p(R_2|M^i_s=s_2)E(t^i|R_2\cap M^i_s=s_2) + p(S|M^i_s=s_2)E(t^i|S\cap M^i_s=s_2) =\\
& \frac{\delta p(R_1)}{\delta p(R_1)+p(R_2) + p(S)}E(t^i|R_1\cap M^i_s=s_2) + \frac{ p(R_2)}{\delta p(R_1)+p(R_2) + p(S)})E(t^i|R_2\cap M^i_s=s_2)  + \\
& \frac{p(S)}{\delta p(R_1)+p(R_2) + p(S)}E(t^i|S\cap M^i_s=s_2)
\end{split}
\end{equation}

from equation \ref{eqt171} we get
\begin{equation} \label{eqt70}
\begin{aligned}
\forall K < i \leq N: E(t^i|\delta=1 \cap m^i_s = s_2) = \mu_R > b
\end{aligned}
\end{equation}

\begin{equation} \label{eqt71}
\begin{aligned}
\forall K < i \leq N: E(t^i|\delta=0 \cap m^i_s = s_2) = \frac{p(R_2)E(t^i|R_2 \cap M^i_s=s_2)}{p(R_2)+p(S)} + \frac{p(S)E(t^i|S \cap M^i_s=s_2)}{p(R_2)+p(S)}
\end{aligned}
\end{equation}
Let $\epsilon' = min(0.5,\frac{(1-q)(b - p_L)}{2})$ and $K=K(\epsilon')$ be the corresponding sample size from Lemma \ref{THM1}:
 \begin{equation} \label{eqt771}
\begin{split}
&\forall K < i \leq N: E(t^i|R_2 \cap M^i_s=s_2) = p(H|R_2 \cap M^i_s=s_2)p_H + p(L|R_2 \cap M^i_s=s_2)p_L \leq \\
&\epsilon' p_H + (1-\epsilon')p_L \leq p_L+\epsilon'
\end{split}
\end{equation}
Lemma \ref{THM2} implies:
 \begin{equation} \label{eqt772}
\begin{aligned}
p_L+\epsilon' < b
\end{aligned}
\end{equation}
 Given $m^i_s = s_2$ implies the first $K$ agents followes the recommendation of the mediator. Hence we get:
\begin{equation} \label{eqt72}
\begin{aligned}
\forall K < i \leq N: E(t^i|S \cap M^i_s=s_2) = E(t^i|a^K = S \cap M^i_s=s_2) < b
\end{aligned}
\end{equation}
Substituting equations \ref{eqt771}, \ref{eqt772} and \ref{eqt72} into equation \ref{eqt71}:
\begin{equation} \label{eqt73}
\begin{split}
&\forall K < i \leq N:\\
&E(t^i|\delta=0 \cap m^i_s = s_2) \leq \frac{p(R_2)(p_L+\epsilon')}{p(R_2)+p(S)} + \frac{p(S)E(t^i|S \cap M^i_s=s_2)}{p(R_2)+p(S)} <\\
&\frac{p(R_2)b}{p(R_2)+p(S)} + \frac{p(S)b}{p(R_2)+p(S)} =b
\end{split}
\end{equation}

For any other $\delta \in (0,1)$ we get:
\begin{equation} \label{eqt74}
\begin{split}
&\forall K < i \leq N: E(t^i|m^i_s=s_2) = \\
&\frac{P(R_1)\delta E(t^i|R_1 \cap m^i_s=s_2)+p(R_2)E(t^i|R_2 \cap m^i_s=s_2) + p(S)E(t^i|S \cap m^i_s=s_2)}{p(R_1)\delta+p(R_2)+p(S)}
\end{split}
\end{equation}
Recall that we assume $q \neq 0,1$ and so $p(R_2) + p(S) > 0$ (to see this note that whenever $p(S) = 0$ all $K$ agents necessarily pulled the risky arm, which implies that $p(R_1) = q$ and $p(R_2) = (1-q)$. Since $q \neq 1$ we get that $p(R_2) > 0$).

Therefore, $E(t^i|m^i_s=s_2)$ must be a continuous as a function of the mediator parameter $\delta$.

From the intermediate value theorem and and inequalities \ref{eqt70} and \ref{eqt73} the desired conclusion follows.

Q.E.D. \\

\subsection{The switching phase}

Suppose an agent knows that the mediator is in the switching phase and, in addition, observes his predecessor pulling the risky arm and receiving a high reward. In that case the agent will prefer the risky arm over the safe arm:

\begin{lemma}\label{THMSTAY1}

 For any $\epsilon, \beta >0$ there exist $K'=K(\epsilon)$ and $ \delta \in[0,1]$ such that for any  $N>K>K'$ and any $N \geq  i >K$, the innkeeper mediator $INKP(N, \beta, \delta, K)$ satisfies
 $E(t^i|r^{i-1}=1 \cap m^i_s = s_2) \geq  b$.
\end{lemma}

\textbf{Proof}:

\begin{equation}\label{eqt31}
\begin{split}
&p(H|r^{i-1}=1 \cap a^{i-1} = R \cap m^i_s = s_2) = \frac{p(H \cap r^{i-1}=1 \cap a^{i-1} = R \cap m^i_s = s_2)}{p(r^{i-1}=1 \cap a^{i-1} = R \cap m^i_s = s_2)} = \\
&\frac{p(m^i_s = s_2\cap a^{i-1} = R)p(H|m^i_s = s_2\cap a^{i-1} = R)p(r^{i-1}=1|H\cap m^i_s = s_2\cap a^{i-1} = R)}{p(m^i_s = s_2\cap a^{i-1} = R)p(r^{i-1}=1|m^i_s = s_2\cap a^{i-1} = R)} =\\
&p(H|m^i_s = s_2 \cap a^{i-1} = R) \frac{p_H}{p(H|m^i_s = s_2\cap a^{i-1} = R)p_H+p(L|m^i_s = s_2\cap a^{i-1} = R)p_L} > \\
&p(H|m^i_s = s_2\cap a^{i-1} = R)\frac{p_H}{p(H|m^i_s = s_2\cap a^{i-1} = R)p_H+p(L|m^i_s = s_2\cap a^{i-1} = R)p_H} = \\
&p(H|m^i_s = s_2\cap a^{i-1} = R)
\end{split}
\end{equation}
Let $\eta = p(H|r^{i-1}=1 \cap a^{i-1} = R \cap m^i_s = s_2)- p(H| a^{i-1} = R \cap m^i_s = s_2)$ Hence from equation \ref{eqt31} we get:
\begin{equation}\label{eqt32}
\begin{aligned}
p(H|r^{i-1}=1 \cap a^{i-1} = R \cap m^i_s = s_2)- p(H| a^{i-1} = R \cap m^i_s = s_2) = \eta > 0
\end{aligned}
\end{equation}

Given $m^i_s = s_2$ implies all previous agents followed the recommendation of the mediator hence we get:
\begin{equation} \label{eqt33}
\begin{split}
&E(t^i|a^{i-1} = R \cap m^i_s = s_2)\geq b \Rightarrow \\
&p(H|a^{i-1} = R \cap m^i_s = s_2)E(t^i|H \cap a^{i-1} = R \cap m^i_s = s_2) +\\
& p(L|a^{i-1} = R \cap m^i_s = s_2)E(t^i|L \cap a^{i-1} = R \cap m^i_s = s_2) \geq b \Rightarrow  \\
&p(H|a^{i-1} = R \cap m^i_s = s_2)p_H+ p(L|a^{i-1} = R \cap m^i_s = s_2)p_L \geq b
\end{split}
\end{equation}

\begin{equation} \label{eqt34}
\begin{split}
&E(t^i|r^{i-1}=1 \cap a^{i-1} = R \cap m^i_s = s_2)= \\
&p(H|r^{i-1}=1 \cap a^{i-1} = R \cap m^i_s = s_2)E(t^i|H \cap r^{i-1}=1 \cap a^{i-1} = R \cap m^i_s = s_2) + \\
&p(L|r^{i-1}=1 \cap a^{i-1} = R \cap m^i_s = s_2)E(t^i|L \cap r^{i-1}=1 \cap a^{i-1} = R \cap m^i_s = s_2) = \\
&p(H|r^{i-1}=1 \cap a^{i-1} = R \cap m^i_s = s_2)p_H + p(L|r^{i-1}=1 \cap a^{i-1} = R \cap m^i_s = s_2)p_L.
\end{split}
\end{equation}

Substituting equation \ref{eqt32} into equation \ref{eqt34}:
\begin{equation} \label{eqt35}
\begin{split}
&E(t^i|r^{i-1}=1 \cap a^{i-1} = R \cap m^i_s = s_2)= \\
&(p(H|a^{i-1} = R \cap m^i_s = s_2)+\eta )p_H +  [1-(p(H|a^{i-1} = R \cap m^i_s = s_2)+\eta)]p_L = \\
&(p(H|a^{i-1} = R\cap m^i_s = s_2)p_H) + (1-p(H|a^{i-1} = R\cap m^i_s = s_2))p_L + \\
&\eta p_H - \eta p_L.
\end{split}
\end{equation}
However, since $(1-p(H|a^{i-1} = R \cap m^i_s = s_2) = p(L|a^{i-1} = R \cap m^i_s = s_2)$ then we can substitute equation \ref{eqt33} in equation \ref{eqt35} to get
\begin{equation} \label{eqt36}
\begin{split}
&E(t^i|r^{i-1}=1 \cap a^{i-1} = R \cap m^i_s = s_2) \geq \\
&b + \eta p_H - \eta p_L  > b
\end{split}
\end{equation}
Q.E.D. \\

\subsection{Success in the switching phase}.

The switching phase is deemed successful once $2K$ switches have been observed. Recall that because we deal with finite populations the switching phase is not guaranteed to be successful. Recall that $X$ counts the number of times during the switching phase that an agent observed his predecessor receive the reward $0$ from the risky arm. Therefore the event that the switching phase is successful can be denoted by $\{X=K\}$.

The following lemmas provide an analysis of the mediator and agent's incentives, conditional on the success of the switching phase.


In the following lemma we make use of the following notation: $d=\{(R_1\cap cf=1) \cup R_2 \cup S\}$. In words, $d$ denotes the event that the mediator moved to the switching phase after the first $K$ agents. The lemma claims that all agents in the population face the same probability of being recommended a switch, conditional on the event that indeed the switching phase was successful. In particular, this probability is equal  $\frac{K}{N}$. This holds even when conditioned on the realized state and the actual event that triggered the switching phase.

\begin{lemma}\label{lem1}
For any of the three cases that trigger switching, $Z\in \{(R_1\cap cf=1),R_2,S\}$, for any agent $1 \leq i \leq N$ and for any event whereby the mediator recommends a switch during the switching phase, $Y_i\in \{(r^{\sigma(i)-1}=0 \cap m^{\sigma(i)}_s = s_2),(r^{\sigma(i)-1}=b\cap m^{\sigma(i)}_s = s_2)\}$:
\begin{enumerate}
\item $p(Y_i|H \cap Z\cap d \cap \{X=K\}) = \frac{K}{N}$
\item $p(Y_i|L \cap Z\cap d \cap \{X=K\}) =\frac{K}{N}$
\item $ p(Y_i|d \cap \{X=K\}) = \frac{K}{N}$
\item $p(Y_i|Z\cap d \cap \{X=K\}) = \frac{K}{N}$  (Follows immediately from (1) and (2)).
\end{enumerate}
\end{lemma}

\textbf{Proof}:
\begin{itemize}
\item case (1) - $Y_i = (r^{\sigma(i)-1}=0 \cap m^{\sigma(i)}_s = s_2)$ :
	\begin{enumerate}
     \item $p(Y_i|H \cap Z\cap d \cap \{X=K\}) = p(x_{\sigma(i)}=1|H \cap Z\cap d \cap \{X=K\})$
     \item $p(Y_i|L \cap Z\cap d \cap \{X=K\}) = p(x_{\sigma(i)}=1|L \cap Z\cap d \cap \{X=K\})$
     \item $p(Y_i|d \cap \{X=K\}) = p(x_{\sigma(i)}=1|d \cap \{X=K\})$
	\end{enumerate}
However, given $X=\Sigma_{i=1}^Nx_i=K$ since $\sigma$ is distributed uniformly we get: $p(x_{\sigma(i)}=1|H \cap Z\cap d \cap \{X=K\}) = p(x_{\sigma(i)}=1|L \cap Z\cap d \cap \{X=K\}) = p(x_{\sigma(i)}=1|d \cap \{X=K\}) = \frac{K}{N}$.

Q.E.D case (1)
\item case (2) - $Y_i = (r^{\sigma(i)-1}=b \cap m^{\sigma(i)}_s = s_2)$ :
\begin{enumerate}
     \item $p(Y_i|H \cap Z\cap d \cap \{X=K\}) = p(x_{\sigma(i)-1}=1|H \cap Z\cap d \cap \{X=K\})$
     \item $p(Y_i|L \cap Z\cap d \cap \{X=K\}) = p(x_{\sigma(i)-1}=1|L \cap Z\cap d \cap \{X=K\})$
     \item $p(Y_i|d \cap \{X=K\}) = p(x_{\sigma(i)-1}=1|d \cap \{X=K\})$
	\end{enumerate}
\footnote{Note that the above equations holds if $\{(R_1\cap cf=1) \cup R_2$\} else (if \{S\}) let
$mx = max (i: 1\leq i\leq N | x_i=1)$, and $mx$ must be switched with $K$. i.e. The first agent in the experiment will observe $b$ and not the last one}
However, given $X=\Sigma_{i=1}^Nx_i=K$ since $\sigma$ is distributed uniformly we get:$p(x_{\sigma(i)-1}=1|H \cap Z\cap d \cap \{X=K\}) = p(x_{\sigma(i)-1}=1|L \cap Z\cap d \cap \{X=K\}) = p(x_{\sigma(i)-1}=1|d \cap \{X=K\}) = \frac{K}{N}$.

Q.E.D case (2)
\end{itemize}

In the following lemma we argue that the expected reward from the risky arm and the safe arm are equal conditional on the following information: (a) A switch from the safe to the risky arm is recommended and a successful switching phase is guaranteed. (b) A switch from the risky to the safe arm is recommended and a successful switching phase is guaranteed:

\begin{lemma}\label{THMdangle}
For any agent after the pre-intervention phase, $i > K$, the following expected rewards are equal:  $$E(t^{\sigma(i)}|r^{\sigma(i)-1}=0 \cap \{m^{\sigma(i)}_s = s_2\} \cap \{X=K\}) = $$
$$E(t^{\sigma(i)}|r^{\sigma(i)-1}=b \cap \{m^{\sigma(i)}_s = s_2\} \cap \{X=K\}) = $$
$$E(t^{\sigma(i)}| d \cap \{X=K\}).$$
\end{lemma}

\textbf{Proof}:

$\forall Z\in \{(R_1\cap cf=1),R_2,S\}, Y_i\in \{(r^{\sigma(i)-1}=0 \cap m^{\sigma(i)}_s = s_2),(r^{\sigma(i)-1}=b\cap m^{\sigma(i)}_s = s_2)\},  1 \leq i \leq N:$\\
By bayes rule:
\begin{equation} \label{eqt141}
\begin{split}
& p(Z|Y_i\cap d \cap \{X=K\}) =
\frac{p(Z|d \cap \{X=K\})p(Y_i|Z\cap d \cap \{X=K\})}{p(Y_i|d \cap \{X=K\})}\\
\end{split}
\end{equation}

Applying Lemma \ref{lem1} into equation \ref{eqt141} We get:
\begin{equation} \label{eqt144}
\begin{split}
&p(Z|Y_i\cap d \cap \{X=K\})  = p(Z|d \cap \{X=K\})\\
\end{split}
\end{equation}
However since $\bigcup_{Z\in \{(R_1\cap cf=1),R_2,S\}}=d$ we get:
\begin{equation} \label{eqt145}
\begin{split}
&E(t^{\sigma(i)}|Y_i\cap d \cap \{X=K\}) = \\
&\Sigma_{Z\in \{(R_1\cap cf=1) , R_2 , S\}}p(Z|Y_i\cap d \cap \{X=K\})
E(t^{\sigma(i)}|Y_i\cap Z \cap d \cap \{X=K\}))
\end{split}
\end{equation}

Applying equation \ref{eqt144} into equation \ref{eqt145} we get:
\begin{equation} \label{eqt146}
\begin{split}
&E(t^{\sigma(i)}|Y_i\cap d \cap \{X=K\}) = \\
&\Sigma_{Z\in \{(R_1\cap cf=1) , R_2 , S\}}p(Z|d \cap \{X=K\})
E(t^{\sigma(i)}|Y_i\cap Z \cap d \cap \{X=K\})\\
\end{split}
\end{equation}


From Lemma \ref{lem1} we get:
\begin{equation} \label{eqt150}
\begin{split}
&E(t^{\sigma(i)}|Y_i \cap Z \cap d \cap \{x=K\}) = p(H|Y_i\cap Z \cap d \cap \{x=K\})p_H + p(L|Y_i\cap Z\cap d \cap \{x=K\})p_L = \\
&\frac{p(H \cap Y_i \cap Z\cap d \cap \{x=K\})p_H}{p(Y_i\cap Z\cap d\cap \{x=K\})} + \frac{p(L \cap Y_i \cap Z\cap d\cap \{x=K\})p_L}{p(Y_i\cap Z\cap d\cap \{x=K\})} =\\
&\frac{p(d\cap Z\cap \{x=K\})p(H|Z\cap d\cap \{x=K\})p(Y_i|H\cap Z\cap d \cap \{x=K\})p_H}{p(d\cap Z\cap \{x=K\})p(Y_i|Z\cap d\cap \{x=K\})} \\
&+ \frac{p(d\cap Z\cap \{x=K\})p(L|Z\cap d\cap \{x=K\})p(Y_i|L\cap Z\cap d \cap \{x=K\})p_L}{p(d\cap Z\cap \{x=K\})p(Y_i|Z\cap d\cap \{x=K\})} = \\
&p(H|Z\cap d \cap \{x=K\})p_H + p(L|Z\cap d \cap \{x=K\})p_L = E(t^{\sigma(i)}|Z\cap d\cap \{x=K\} )
\end{split}
\end{equation}

Applying equation  \ref{eqt150} into equation \ref{eqt146} we get:
\begin{equation} \label{eqt153}
\begin{split}
&E(t^{\sigma(i)}|Y_i\cap d \cap \{X=K\}) = \\
&\Sigma_{Z\in \{(R_1\cap cf=1) , R_2 , S\}}p(Z|d \cap \{X=K\})
E(t^{\sigma(i)}| Z \cap d \cap \{X=K\}) =\\
&E(t^{\sigma(i)}| d \cap \{X=K\})
\end{split}
\end{equation}
However, since $(Y_i \cap (cf=0)) = \emptyset$ we get:
\begin{equation} \label{eqt1153}
\begin{split}
   E(t^{\sigma(i)}|Y_i\cap d \cap \{X=K\}) = E(t^{\sigma(i)}|Y_i \cap \{X=K\})
\end{split}
\end{equation}
From equations \ref{eqt153} and \ref{eqt1153} we get:
\begin{equation} \label{eqt1154}
\begin{split}
&E(t^{\sigma(i)}|r^{\sigma(i)-1}=0 \cap \{m^{\sigma(i)}_s = s_2\} \cap \{X=K\})\\
& = E(t^{\sigma(i)}|r^{\sigma(i)-1}=b \cap \{m^{\sigma(i)}_s = s_2\} \cap \{X=K\})\\
& =  E(t^i| d \cap \{X=K\})
\end{split}
\end{equation}
Q.E.D \\

\subsection{The actual length of the switching phase}

The length of the switching phase is a random variable and depends on the number of times the risky arm is pulled and a low reward is realized. Given $n$ attempts of pulling the risky arm, let $Y$ be  the number of occurrences of low rewards. Note that $Y$ is a binomial random variable with distribution $B(n,1-p_{\omega})$, at state $\omega$. By choosing $n$ large enough we can provide an upper bound on the probability of the event that sufficiently many switches occur:

\begin{lemma}\label{THM11}
Assume $Y \sim B(n,1-p_H)$ and $X \sim B(n, 1-p_L)$. Then $\forall K>0$, $\beta>0$ there exists $n'$ such that $\forall n>n'$ :
 \begin{enumerate}
   \item $p(Y<K) \leq \frac{\beta}{4K}$
   \item $p(X<K) \leq \frac{\beta}{4K}$
 \end{enumerate}
\end{lemma}

\textbf{Proof}:

\begin{equation}\label{eqth011}
\begin{aligned}
Prob(Y < K) = Prob(E(Y)- Y > E(Y)-K) \le \\
\le Prob(E(Y)- Y > E(Y)-K)) + Prob(Y-E(Y) > E(Y)-K) = \\
= Prob(|E(Y)- X| > E(Y)-K).
\end{aligned}
\end{equation}

Applying Chebyshev inequality:
\begin{equation}\label{eqth012}
\begin{aligned}
Prob(|E(Y)- Y| > E(Y)-K) \le  \frac {Var(Y)}{(K-E(Y))^{2}} =
\frac {np_H(1-p_H)}{(K-n(1-p_H))^{2}}
\end{aligned}
\end{equation}

It suffices to prove the first part as the second part follows the same arguments, and since $P_H>p_L$ same $n'$ will hold.

\begin{equation}\label{eqth121}
\begin{aligned}
\frac {np_H(1-p_H)}{(K-n(1-p_H))^{2}} \leq \frac{\beta}{4K} \Leftrightarrow\\
np_H(1-p_H) \leq (K-n(1-p_H))^{2}\frac{\beta}{4K} \Leftrightarrow\\
\frac{np_H(1-p_H)}{\frac{\beta}{4K}} \leq K^2 - 2k(1-p_H)n + (1-p_H)^2n^2 \Leftrightarrow\\
 (1-p_H)^2n^2 - (1-p_H)(2K+\frac{p_H}{\frac{\beta}{4K}})n + K^2 \geq 0 \Leftrightarrow \\
 n \geq \frac{(1-p_H)(2K+\frac{p_H}{\frac{\beta}{4K}})+\sqrt{[(1-p_H)(2K+\frac{p_H}{\frac{\beta}{4K}})]^2-4(1-p_H)^2K^2}}{2(1-p_H)^2}
\end{aligned}
\end{equation}
Let $n' = \frac{(1-p_H)(2K+\frac{p_H}{\frac{\beta}{4K}})+\sqrt{[(1-p_H)(2K+\frac{p_H}{\frac{\beta}{4K}})]^2-4(1-p_H)^2K^2}}{2(1-p_H)^2}$ the proof followed from equations \ref{eqth121} and \ref{eqth012}. \\

QED \\

\subsection{The introduction of transfers}

In Lemma \ref{THMdangle} we proved that in the hypothetical case that $2K$ switches is guaranteed the agents will be fully indifference between the two arms case they get the switch signal. In lemma \ref{THM11} we prove we can determine the time needed to ensure $2K$ switches with high enough probability. In the following lemma we will prove that the impossibility to guarantee $2K$ switches can be overcome by using transfers.  The magnitude of these transfers is determined in lemma \ref{THM11} and can be arbitrarily low at the expense of the proportion of agents to enjoy the exploitation phase.

\begin{lemma}\label{1THMdangle}
Given the model parameters $p_H, p_L,q$ and $b$,
the needed asymptotic optimality factor $\epsilon$, then $\forall \beta>0 \ \ and \ \ K>0 \ \  \exists N'|\forall N\geq N' \ $ and $\delta \in [0,1]$ such that the innkeeper mediator $INKP( N, \beta, \delta, K)$ satisfies:
\begin{enumerate}
\item $\forall 1 < i \leq N: E(t^{\sigma(i)}|r^{\sigma(i)-1}=0\cap \{m^{\sigma(i)}_s = s_2\}) \leq b + \frac{\beta}{2K}$
\item $\forall 1 < i  \leq N:E(t^{\sigma(i)}|r^{\sigma(i)-1}=b\cap \{m^{\sigma(i)}_s = s_2\}) + \frac{\beta}{2K} \geq b$
\end{enumerate}
\end{lemma}

\textbf{Proof}:\\
Let us mark $\forall 1 < i \leq N$: $Y^1_i = \{(r^{\sigma(i)-1}=0 \cap m^{\sigma(i)}_s = s_2 \}$ and $Y^2_i = \{r^{\sigma(i)-1}=b\cap m^{\sigma(i)}_s = s_2)\}$ . Recall from Lemma \ref{THM11} that $n' = \frac{(1-p_H)(2K+\frac{p_H}{\frac{\beta}{4K}})+\sqrt{[(1-p_H)(2K+\frac{p_H}{\frac{\beta}{4K}})]^2-4(1-p_H)^2K^2}}{2(1-p_H)^2}$ and let $N' = n'+K$.\\
Hence from Lemma \ref{THM11} we get:
\begin{equation} \label{eqt1451}
\begin{split}
&p(x=K|Y^1_i\cap d) \geq  1-\frac{\beta}{4K} ; \\
&p(x=K|Y^2_i\cap d) \geq  1-\frac{\beta}{4K}
\end{split}
\end{equation}

Hence we get:
\begin{equation} \label{eqt15312}
\begin{split}
&E(t^{\sigma(i)}|Y^1_i\cap d)= \\
&p(x=K|Y^1_i\cap d)E(t^{\sigma(i)}|Y^1_i\cap d \cap x=K)+
p(x<K|Y^1_i\cap d)E(t^{\sigma(i)}|Y^1_i\cap d \cap x<K) \leq \\
&1E(t^{\sigma(i)}|Y^1_i\cap d \cap x=K)+ \frac{\beta}{4K}E(t^{\sigma(i)}|Y^1_i\cap d \cap x<K)) \leq \\
&E(t^{\sigma(i)}|Y^1_i\cap d \cap x=K)+ \frac{\beta}{4K}
\end{split}
\end{equation}
From Lemma \ref{THMdangle} and equation \ref{eqt15312} we get:
\begin{equation} \label{eqt1531}
\begin{split}
&E(t^{\sigma(i)}|Y^1_i\cap d) \leq E(t^{\sigma(i)}| d \cap x=K)+ \frac{\beta}{4K}  =\\
&p(x=K|d)E(t^{\sigma(i)}| d \cap x=K) +
(1-p(x=K|d))E(t^{\sigma(i)}| d \cap x=K) +\\
&p(x<K|d)E(t^{\sigma(i)}| d \cap x<K)- p(x<K|d)E(t^{\sigma(i)}| d \cap x<K) + \frac{\beta}{4K} \leq \\
&[ p(x=K|d)E(t^{\sigma(i)}| d \cap x=K)+
p(x<K|d)E(t^{\sigma(i)}| d \cap x<K)]+ \\
&(1-p(x=K|d))E(t^{\sigma(i)}| d \cap x=K) + \frac{\beta}{4K} \leq \\
&E(t^{\sigma(i)}|d) +  \frac{\beta}{4K}E(t^{\sigma(i)}|d \cap x=K) + \frac{\beta}{4K} \leq b + 2\frac{\beta}{4K}= b + \frac{\beta}{2K}
\end{split}
\end{equation}
Since $(Y^1_i \cap (cf=0)) = \emptyset$ we get:
\begin{equation} \label{eqt11451}
\begin{split}
&E(t^{\sigma(i)}|Y^1_i) = E(t^{\sigma(i)}|Y^1_i\cap d)
\end{split}
\end{equation}
From equations \ref{eqt1531} and \ref{eqt11451} we get:
\begin{equation} \label{eqt21451}
\begin{split}
&E(t^{\sigma(i)}|Y^1_i) \leq b + \frac{\beta}{2K}
\end{split}
\end{equation}
Q.E.D(1)

On the other hand from equation \ref{eqt1451} we also get:

\begin{equation} \label{eqt25312}
\begin{split}
&E(t^{\sigma(i)}|Y^2_i\cap d)= \\
&p(x=K|Y^2_i\cap d)E(t^{\sigma(i)}|Y^2_i\cap d \cap x=K)+
p(x<K|Y^2_i\cap d)E(t^{\sigma(i)}|Y^2_i\cap d \cap x<K) \geq \\
&(1-\frac{\beta}{4K})E(t^{\sigma(i)}|Y^2_i\cap d \cap x=K) \geq
E(t^{\sigma(i)}|Y^2_i\cap d \cap x=K)- \frac{\beta}{4K}
\end{split}
\end{equation}
From Lemma \ref{THMdangle} and equation \ref{eqt25312} we get:
\begin{equation} \label{eqt2531}
\begin{split}
&E(t^{\sigma(i)}|Y^2_i\cap d) \geq E(t^{\sigma(i)}| d \cap x=K)- \frac{\beta}{4K}  =\\
&p(x=K|d)E(t^{\sigma(i)}| d \cap x=K) +
(1-p(x=K|d))E(t^{\sigma(i)}| d \cap x=K) +\\
&p(x<K|d)E(t^{\sigma(i)}| d \cap x<K)- p(x<K|d)E(t^{\sigma(i)}| d \cap x<K) - \frac{\beta}{4K} \geq \\
&[ p(x=K|d)E(t^{\sigma(i)}| d \cap x=K)+
p(x<K|d)E(t^{\sigma(i)}| d \cap x<K)]- \\
& -p(x<K|d)E(t^{\sigma(i)}| d \cap x<K) - \frac{\beta}{4K} \geq \\
&E(t^{\sigma(i)}|d) -  \frac{\beta}{4K}E(t^{\sigma(i)}| d \cap x<K) - \frac{\beta}{4K} \geq b - 2\frac{\beta}{4K}= b - \frac{\beta}{2K}
\end{split}
\end{equation}
Since $(Y^2_i \cap (cf=0)) = \emptyset$ we get:
\begin{equation} \label{eqtb21451}
\begin{split}
&E(t^{\sigma(i)}|Y^2_i) = E(t^{\sigma(i)}|Y^2_i\cap d)
\end{split}
\end{equation}
From equations \ref{eqt2531} and \ref{eqtb21451} we get:
\begin{equation} \label{eqt31451}
\begin{split}
&E(t^{\sigma(i)}|Y^2_i)  + \frac{\beta}{2K} \geq  b
\end{split}
\end{equation}
Q.E.D(2)

\subsection{Proof of main theorem}

Recall the main Theorem: \\
\textbf {Theorem \ref{THEOREM1}:} For any $\epsilon, \beta>0$ there exist $N' \ge K>0$ and $\delta \in [0,1]$ such that the innkeeper mediator $INKP( N, \beta, \delta, K)$ is incentive compatible, $\beta$-budget balanced and $\epsilon$-optimal for all $N\geq N'\ $. \\
\textbf{Proof of Theorem \ref{THEOREM1} }:

Given $\epsilon, \beta>0$, we will first offer the appropriate $K$ and $N'$ and we will use Lemma \ref{THMDELTA} for the existence of an appropriate $\delta$. Afterwards We will break down the proof into three propositions which correspond to the three claims stated in the theorem: incentive compatibility, $\epsilon$-optimality and $\beta$-budget balanced.\\

\noindent{\bf Finding an appropriate parameter $K$:}\\
Given $\epsilon, \beta>0$ we will demand that $K$ will be big enough both to satisfy all the relevant Lemmas needed in our proof and to satisfy the demands for $\epsilon$-optimal.
\begin{itemize}
   \item To satisfy conditions for Lemma \ref{THM1} with $\frac{\epsilon}{4}$ (needed for $\epsilon$-optimal): Set $K_{1H} = \frac{4p_H(1-p_H)}{\frac{\epsilon}{4}(p_H-p_L)^2}$, $K_{1L} = \frac{4p_L(1-p_L)}{\frac{\epsilon}{4}(p_H-p_L)^2}$ and
$K_1 = max(K_{1H},K_{1L})$.
\item To satisfy conditions for Lemma \ref{THM2}:  Set $K_{2H} = \frac{4p_H(1-p_H)}{min(0.5,\frac{(1-q)(b - p_L)}{2})(p_H-p_L)^2}$, \\ $K_{2L} = \frac{4p_L(1-p_L)}{min(0.5,\frac{(1-q)(b - p_L)}{2})(p_H-p_L)^2}$ and
$K_2 = max(K_{2H},K_{2L})$.
\item K that satisfies the conditions of Lemmas \ref{THM1} and \ref{THM2} satisfies the conditions of lemmas \ref{THMDELTA}, \ref{THMSTAY1}, \ref{lem1}, \ref{THMdangle}, \ref{THM11} and \ref{1THMdangle}.
\end{itemize}
Hence the appropriate $K$ to satisfy all our needed conditions will be $K= max(K_1, K_2)$\\

\noindent{\bf Finding an appropriate parameter $N'$:}

After finding an appropriate parameter $K$ we will now turn to find an appropriate parameter $N'$ big enough both to satisfy all the relevant Lemmas needed in our proof and to satisfy the demands for $\epsilon$-optimal (note that $N'$ depends on $K$ and therefore we had to find an appropriate $K$ first).

Let $ \hat{n} = \frac{(1-p_H)(2K+\frac{p_H}{min(\frac{\beta}{4K}, \frac{\epsilon}{4})})+\sqrt{[(1-p_H)(2K+\frac{p_H}{min(\frac{\beta}{4K}, \frac{\epsilon}{4})})]^2-4(1-p_H)^2K^2}}{2(1-p_H)^2}$

\begin{itemize}
   \item To satisfy conditions for Lemmas \ref{THM11} and \ref{1THMdangle} we will demand: \\ $N'_1 = K + \hat{n}$
   \item To satisfy $\epsilon$-optimal we will demand $N'_2 = \frac{2}{\epsilon} [K + \hat{n}] > N'_1$
\end{itemize}
Hence the appropriate $N'$ to satisfy all our needed conditions will be $N' = \frac{2}{\epsilon} [K + \hat{n}]$\\

\noindent{\bf existence of appropriate parameter $\delta$:}\\
Given $\epsilon, \beta>0$, and finding appropriate $K$ and $N'$. From Lemma \ref{THMDELTA} there exists $ \delta \in[0,1]$ such that for any  $N>N'$ and any $N \geq  i >K$, the innkeeper mediator $INKP(N, \beta, \delta, K)$ satisfies  $E(t^i|M^i_s=s_2)=b$. \\

\noindent{\bf Proof break down to three propositions :}\\
After we offered the appropriate $K$ and $N'$ and used Lemma \ref{THMDELTA} for the existence of an appropriate $\delta$, we break down the proof into three propositions which correspond to the three claims stated in the theorem: incentive compatibility, $\epsilon$-optimality and $\beta$-budget balanced.\\

\noindent{\bf Incentive compatibility:}
\begin{proposition}\label{prop1}
   For every $N>N'$, the innkeeper mediator $INKP( N, \beta, \delta, K)$ is incentive compatible.
\end{proposition}

\textbf{Proof}:

  The first agent will get the message $(R,s_1,0)$ and will optimally comply since $qp_H+(1-q)p_L > b$.
We now consider 4 cases: agents getting $m^i_s = s_1 (pre\_intervention)$ in the message, agents getting $m^i_s = s_2 (switching)$ in the message, agents getting $m^i_s =s_3 (exploit)$ in the message and agents getting $m^i_s =s_4 (deviation)$ in the message

\begin{enumerate}
\item
Consider an agent getting $m^i_s = s_1$ in the message. Since $m^i_s \neq s_4$  all other agents follow the recommendation of the mediator. In this case the agent can deduce he is among the first $K$ arriving agents. However, since for those agents the mediator calculate and recommend the arm which is the best response according to the agent's knowledge he will comply and pull the recommended arm.
\item
Consider an agent getting $m^i_s = s_2$ in the message.  Since $m^i_s \neq s_4$  all other agents follow the recommendation of the mediator. There are 3 possible cases:
\begin{enumerate}
\item The agent observe his predecessor pulled the risky arm and rewarded $1$. In this case according to the mediator he will get the message $\tilde M^{\sigma(i)} = (R,s_2,0)$. According to Lemma \ref{THMSTAY1} $E(t^{\sigma(i)}| r^{\sigma(i)-1}=1 \cap \{m^{\sigma(i)}_s=s_2\}) \geq b$ and therefore his best response will be to pull the risky arm and therefore he will comply.
\item The agent observe his predecessor pulled the risky arm and rewarded $0$. In this case according to the mediator he will get the message $\tilde M^{\sigma(i)} = (S,s_2,\frac{\beta}{2K})$. According to Lemma \ref{1THMdangle} $E(t^{\sigma(i)}|r^{\sigma(i)-1}=0\cap \{m^{\sigma(i)}_s = s_2\}) \leq b + \frac{\beta}{2K}$ and therefore his best response will be to pull the safe arm and get extra payment of $\frac{\beta}{2K}$ for doing so. Therefore he will comply.
\item The agent observe his predecessor pulled the safe arm and rewarded $b$. In this case according to the mediator he will get the message $\tilde M^{\sigma(i)} = (R,s_2,\frac{\beta}{2K})$. According to Lemma \ref{1THMdangle} $E(t^{\sigma(i)}|r^{\sigma(i)-1}=b\cap \{m^{\sigma(i)}_s = s_2\}) + \frac{\beta}{2K} \geq b$ and therefore his best response will be to pull the risky arm and get extra payment of $\frac{\beta}{2K}$ for doing so. Therefore he will comply.
\end{enumerate}
\item Consider an agent getting $m^i_s =s_3$ in the message.   Since $m^i_s \neq s_4$  all other agents follow the recommendation of the mediator. Note that $s_3$ is send only in cases the mediator observe enough samples of the risky arm and decide with high enough probability if the state of the risky arm is $H$ or $L$. If it determine the state is $H$ it recommend to pull the risky arm and in such case $\forall j: K+1 <j < N:E(t^j|\frac{\Sigma_{i=1}^{K}t^i}{K}\geq \frac{p_H+p_L}{2})\geq qp_H+(1-q)p_L \geq b$ and therefore the agent will comply. If the mediator determine the state  is $L$ it recommend to pull the safe arm and according to Lemma \ref{THM2} $\forall j| K+1 \leq j \leq N: E(t^j|\frac{\Sigma_{i=1}^{K}t^i}{K}< \frac{p_H+p_L}{2}) < b$ and therefore the agent will comply.
\item Consider an agent getting $m^i_s =s_4$ in the message. In that case let's consider 2 sub-cases:
\begin{itemize}
   \item $exploit\_flag = true$: in that case the mediator already determined before the deviation occur with high enough probability if the state of the risky arm is $H$ or $L$. And therefore the same argument as in the case of $s_3$ holds and the best response of the agent is to comply and pull the recommended arm.
   \item $ exploit\_flag = false$: since in this case  the mediator calculate and recommend the arm which is the best response according to the agent knowledge the best response is to comply and pull the recommended  arm.
\end{itemize}
Since in both cases the best response is to comply and pull the recommended arm we get that whenever an agent getting $s_4$ in the message the best response is to pull the recommended arm.
\end{enumerate}
Q.E.D\\

\noindent{\bf $\epsilon$-optimal:}
\begin{proposition}\label{prop2}
   For every $N>N'$, the innkeeper mediator $INKP( N, \beta, \delta, K)$ is $\epsilon$-optimal.
\end{proposition}
\textbf{Proof}:

Since $K$ satisfies the conditions of Lemma \ref{THM1} we get:
\begin{equation} \label{eqt2000}
\begin{split}
\forall j: K+1 \leq j \leq N: p(\{M^j_a = R\}|H \cap \{M^j_s = s_3\}) \geq 1-\frac{\epsilon}{4}
\end{split}
\end{equation}
And
\begin{equation} \label{eqt2001}
\begin{split}
\forall j: K+1 \leq j \leq N: p(\{M^j_a = S\}|L \cap \{M^j_s = s_3\}) \geq 1-\frac{\epsilon}{4}
\end{split}
\end{equation}
Recall  $x_i = 1$ if $(r^{i-1}=0 \cap \{m^i_s = s_2\})$, $x_i = 0$ elsewhere.  $X=\Sigma_{i=1}^nx_i$  (the number of switching from the risky arm to the safe arm during the switching phase).

From Lemma \ref{THM11} $\exists n$ (we will mark it as $\hat{n}$) such that: \footnote{Let $ \hat{n} = \frac{(1-p_H)(2K+\frac{p_H}{min(\frac{\beta}{4K}, \frac{\epsilon}{4})})+\sqrt{[(1-p_H)(2K+\frac{p_H}{min(\frac{\beta}{4K}, \frac{\epsilon}{4})})]^2-4(1-p_H)^2K^2}}{2(1-p_H)^2}$}
\begin{equation} \label{eqt2002}
\begin{split}
p(X<K|H) \leq min(\frac{\beta}{4K},\frac{\epsilon}{4})
\end{split}
\end{equation}
\begin{equation} \label{eqtb2002}
\begin{split}
p(X<K|L) \leq min(\frac{\beta}{4K},\frac{\epsilon}{4})
\end{split}
\end{equation}

By taking $N' = \frac{2}{\epsilon} [K + \hat{n}]$ we get that:
\begin{equation} \label{eqt2003}
\begin{split}
&\forall 0 \leq i \leq N': p(\{M^i_s = s_3\} | \{X=K\}) \geq \frac{N'-(K+\hat{n})}{N'} = \frac{(\frac{2}{\epsilon}-1)(K+\hat{n})}{\frac{2}{\epsilon}(K+\hat{n})}=(1- \frac{\epsilon}{2})
\end{split}
\end{equation}
From equations \ref{eqt2000}, \ref{eqt2002}, \ref{eqt2003} we get:
\begin{equation} \label{eqt2004}
\begin{split}
&\forall 0 \leq i \leq N': p(\{M^{\sigma(i)}_a = R\} | H)  = \\
&p(\{X=K\}| H)p(\{M^{\sigma(i)}_a = R\}| \{X=K\} \cap H) + \\
&p(\{X<K\}| H)p(\{M^{\sigma(i)}_a = R\}| \{X<K\} \cap H)\geq \\
&p(\{X=K\}| H)p(\{M^{\sigma(i)}_a = R\}| \{X=K\} \cap H) =\\
&p(\{X=K\}| H)p(\{M^{\sigma(i)}_s = s_3 \}| \{X=K\} \cap H)p(\{M^{\sigma(i)}_a = R \}|H \cap  \{M^{\sigma(i)}_s = s_3\} \cap \{X=K)\} + \\
& p(\{X=K\}| H)p(\{M^{\sigma(i)}_s \neq s_3 \}| \{X=K\} \cap H)p(\{M^{\sigma(i)}_a = R \}|H \cap  \{M^{\sigma(i)}_s \neq s_3\} \cap \{X=K)\}\geq \\
&p(\{X=K\}| H)p(\{M^{\sigma(i)}_s = s_3 \}| \{X=K\} \cap H)p(\{M^{\sigma(i)}_a = R \}|H \cap  \{M^{\sigma(i)}_s = s_3\} \cap \{X=K)\}  \geq \\
&(1-\frac{\epsilon}{4})(1- \frac{\epsilon}{2})(1-\frac{\epsilon}{4}) \geq \\
&(1- \frac{\epsilon}{2})(1- \frac{\epsilon}{2}) \geq \\
&1-\epsilon
\end{split}
\end{equation}
From equations \ref{eqt2001}, \ref{eqtb2002}, \ref{eqt2003} we get:
\begin{equation} \label{eqt2005}
\begin{split}
&\forall 0 \leq i \leq N':  p(\{M^{\sigma(i)}_a = S\} | L)  \geq \\
& p(\{X=K\}| L)p(\{M^{\sigma(i)}_s = s_3\} | \{X=K\} \cap L)p(\{M^{\sigma(i)}_a = S\} |L \cap  \{M^{\sigma(i)}_s = s_3 \}\cap \{X=K\} \geq \\
&(1-\frac{\epsilon}{4})(1- \frac{\epsilon}{2})(1-\frac{\epsilon}{4}) \geq \\
&(1- \frac{\epsilon}{2})(1- \frac{\epsilon}{2}) \geq \\
&1-\epsilon
\end{split}
\end{equation}
From equations \ref{eqt2004} , \ref{eqt2005} we get that in equilibrium: If $H$ then $E(\frac{\Sigma_{i=1}^Nr^i}{N}) \geq (1-\epsilon)p_H$ else $E(\frac{\Sigma_{i=1}^Nr^i}{N}) \geq (1-\epsilon)b$  \\
Q.E.D.

\noindent{\bf $\beta$-budget balanced:}
\begin{proposition}\label{prop3}
   For every $N>N'$, the innkeeper mediator $INKP( N, \beta, \delta, K)$ is $\beta$-budget balanced.
\end{proposition}
\textbf{Proof}:

 Recall  $x_i = 1$ if $(r^{i-1}=0 \cap \{m^i_s = s_2\})$, $x_i = 0$ elsewhere. Let $y_i = 1$ if $(r^{i-1}=b \cap \{m^i_s = s_2\})$, $y_i = 0$ elsewhere. Note $\forall i: x_i y_i=0$.
according to the mediator a payment of $\frac{\beta}{2K}$ is offered only in case of $x_i = 1$ or $y_i = 1$. Hence we get:
\begin{equation} \label{eqt2006}
\begin{split}
&\Sigma_{i=1}^N m^i_p =\frac{\beta}{2K} \Sigma_{i=1}^Nx_i + \frac{\beta}{2K}\Sigma_{i=1}^Ny_i \leq \frac{\beta}{2K}K + \frac{\beta}{2K}K =\beta
\end{split}
\end{equation}

Q.E.D.\\

The proof of Theorem \ref{THEOREM1} follows from Propositions \ref{prop1}, \ref{prop2} and \ref{prop3}.\\
Q.E.D

\bibliographystyle{abbrvnat}
\bibliography{bibnew}
\end{document}